\documentclass[lettersize,journal]{IEEEtran}
\usepackage{amsmath,amsfonts}
\usepackage{algorithmic}
\usepackage{algorithm}
\usepackage{array}
\usepackage{cite}
\usepackage{graphicx}
\usepackage[hidelinks]{hyperref}
\usepackage[caption=false,font=normalsize,labelfont=sf,textfont=sf]{subfig}
\usepackage{stfloats}
\usepackage{textcomp}
\usepackage{url}
\usepackage{verbatim}
\usepackage[switch]{lineno}
\usepackage{aas_macros}

\begin{document}

%\linenumbers

\title{Thermal and Electrical Properties of Prototype Readout Components for CMB-S4}

%\author{%IEEE Publication Technology,~\IEEEmembership{Staff,~IEEE,}
        % <-this % stops a space
%Wilber Dominguez, University of New Mexico \\
%Darcy Barron, University of New Mexico \\
%Zeeshan Ahmed, SLAC National Accelerator Laboratory \\
%Amy Bender, Argonne National Laboratory \\
%Sandra Diez, National Institute of Standards and Technology (NIST) \& Colorado School of Mines\\
%Malcolm Durkin, University of Colorado \\
%Tristan Eggenberger, University of New Mexico \\
%Gunther Haller, SLAC National Accelerator Laboratory \\
%Shawn Henderson, SLAC National Accelerator Laboratory\\
%Katherine Hewey, University of Arizona \\
%Johannes Hubmayr, National Institute of Standards and Technology (NIST) \\
%Christopher Rooney, National Institute of Standards and Technology (NIST) \\
%Robinjeet Singh, National Institute of Standards and Technology (NIST) \& University of Colorado\\
%Michael Vissers, National Institute of Standards and Technology (NIST)

%\hfill \break

%For the CMB-S4 Collaboration
        
%\thanks{This paper was produced by the IEEE Publication Technology Group. They are in Piscataway, NJ.}% <-this % stops a space
%\thanks{Manuscript received April 19, 2021; revised August 16, 2021.}
%}

\author{%
Wilber Dominguez$^{1}$,
Darcy R. Barron$^{1}$,
Zeeshan Ahmed$^{2}$,
Amy N. Bender$^{3}$,
Sandra Diez$^{4,5}$,
Malcolm Durkin$^{6}$,
Tristan A. Eggenberger$^{1}$,
Gunther Haller$^{2}$,
Shawn W. Henderson$^{2}$,
Katherine Hewey$^{1,7}$,
Johannes Hubmayr$^{4}$,
Christopher Rooney$^{4}$,
Robinjeet Singh$^{4,6}$,
Michael Vissers$^{4}$ \\
%\leavevmode\\
\hfill \break
$^{1}$University of New Mexico \\
$^{2}$SLAC National Accelerator Laboratory \\
$^{3}$Argonne National Laboratory \\
$^{4}$National Institute of Standards and Technology (NIST) \\
$^{5}$Colorado School of Mines \\
$^{6}$University of Colorado \\
$^{7}$University of Arizona \\
%\leavevmode\\

\hfill \break

For the CMB-S4 Collaboration
}



\maketitle

\begin{abstract}
CMB-S4 is the fourth-generation ground-based cosmic microwave background project, designed to probe the early universe and cosmic inflation. CMB-S4 would achieve its science goals in part by dramatically increasing the number of transition edge sensor (TES) bolometer detectors on the sky. The detector readout system for CMB-S4 is time-division multiplexing (TDM) with a two-stage Superconducting Quantum Interference Device (SQUID) system. To accommodate the large increase in detectors, the size of our camera increases, placing physical constraints on the readout, its wiring, and its power dissipation. Therefore, to optimize readout performance, we need to balance competing design considerations such as thermal load and bandwidth. We present results characterizing the thermal and electrical performance of prototype components, including wiring and SQUID arrays for CMB-S4,  and discuss the impact on overall system performance.
\end{abstract}


\section{Introduction}

The Cosmic Microwave Background (CMB) provides a snapshot of the universe as it existed 
around the time of recombination about 380,000 years after the Big Bang, and has imprinted 
in it information on the conditions of our early Universe. A primary goal of CMB-S4 is to 
observe the B-mode polarization in the CMB generated from gravitational waves from 
inflation. To achieve this goal, a significant increase in sensitivity is required, that must be 
matched with strict control and understanding of systematic effects.  CMB-S4 would achieve 
this with 500,000 transition-edge sensor detectors read out with time-division multiplexing (TDM) \cite{CMBS4_forecasting_2022}.

TDM is a technique that reads out multiple detectors with a single readout chain by rapidly cycling through each detector in time. This technology was chosen for CMB-S4 because it is an established technology that can meet CMB-S4's stringent noise requirements (including white noise and low-frequency noise). It has also demonstrated excellent noise performance and high yield in successful CMB instruments, starting with BICEP3 in 2015 \cite{BKXV_2022}.
The CMB-S4 TDM readout system design (described in detail in \cite{barron2022}) incorporates several low-risk technological advancements to improve the multiplexing factor and therefore reduce cost and integration complexity. These improvements, developed at NIST Boulder, include a faster SQUID Series Array (SSA) design with a
shunt resistor to further increase bandwidth, a new first stage SQUID amplifier (SQ1) design
with higher input mutual inductance, and a two-level switching
architecture that substantially reduces cryogenic wire counts \cite{Doriese2016, Durkin2021, Smith2021, Zeng2013, Dawson2019}.
%Developed at NIST Boulder for improving the readout of X-ray TES detectors, these improvements include a new faster SQUID Series Array (SSA) design with a shunt resistor to further increase bandwidth, a new SQ1 design with higher input mutual inductance, and a two-level switching architecture which substantially reduces cryogenic wire counts \cite{Doriese2016, Durkin2021, Smith2021, Zeng2013, Dawson2019}.
%Cannot find Smith and Zeng
%has TES detectors’ current signals amplified by a SQUID (SQ1) and shunted by a Josephson junction switch at millikelvin temperatures. Many SQ1s are connected to a second-stage 
%SQUID Series Array (SSA) amplifier. This amplified signal is then digitized by room temperature electronics. The cycling of detectors is conducted by activating chip select (CS) and row select (RS) switches. The integration of the CS further decreases the total connections while increasing the number of rows that are read out \cite{durkin}.

\section{Motivation}
Achieving the sensitivity goals of CMB-S4 will require a precise understanding of the instrument 
performance, including the readout electronics. The achievable multiplexing factor for previous implementations of TDM readout for CMB experiments were limited in part by system bandwidth. To improve the multiplexing factor for CMB-S4 to the goal of 80, the readout design will need to demonstrate greater bandwidth compared to previous designs. 
An end-to-end noise and bandwidth model of the TDM readout system for CMB-S4 is described in \cite{goldfinger2024}.  
This work focuses on the portion of the TDM readout that includes the second-stage SSA amplification and associated cold wiring, aiming to understand the thermal and electrical properties and how these parameters affect the integrated readout system performance.

The CMB-S4 readout design requires a significant increase in the quantity and length of cryogenic cabling compared to already fielded CMB instruments.
This includes the twisted-pair cables that connect the room temperature electronics to the cold readout electronics. These must be designed so that the total thermal power on the 50K and 4K stages is within the thermal budget set by cryocooler performance and other major contributors including structural support.  These cables from room temperature must have conductors with low thermal conductivity like manganin, with resulting high electrical resistivity, which can limit system bandwidth.

An additional consideration in the design of these cables is the overall system design for shielding against radio frequency (RF) interference, as cables that are run outside of an RF shielded space must have their own shielding.
Due to other design considerations, the space from room temperature down to the 4K stage may not be shielded, and so these cryogenic cables would need to be shielded, further constraining their achievable thermal and electrical properties.
%The system design for shielding against RF interference is under development, and certain runs of cryogenic cabling may need to be shielded, affecting thermal conductivity and electrical properties. 
%Cable capacitance is affected by cable design and must be considered along with the length of cabling as potential limiting factors for the bandwidth of the readout system and the achievable multiplexing factor. 
Finally, cable capacitance and the length of cabling are potential limiting factors in the bandwidth and achievable multiplexing factor of the readout system

% This SSA motivation was good text but just too much detail for right here

Design and configuration of the SSA amplifier is another aspect of the readout design that has an impact on properties of the system including noise and bandwidth. 
TDM for CMB bolometer readout has 
historically chosen a configuration of the SSA that maximized gain in order to minimize the contribution to readout noise, at some cost to bandwidth. For 
CMB-S4, we are studying the trade-offs to understand 
the best configuration for overall system performance and achievable multiplexing factor.

\section{Measurements}

\subsection{Thermal Tests on Prototype CMB-S4 Cold Readout Wiring}

The thermal properties of a CMB-S4 prototype readout cable were characterized and compared with expectations, including dedicated tests of the thermal conductivity of a wire sample taken from the cable. The entire cable has 100 wires packaged as 50 twisted pair cables. Each twisted pair has two manganin wires with fluorinated ethylene propylene (FEP) insulation, with stainless steel (SS 304) shielding over the twisted pair of wires, and another layer of FEP insulation around the entire bundle. 

The thermal conductivity of Manganin and SS 304 is well documented at cryogenic temperatures \cite{duthil2015}. The thermal conductivity of FEP at cryogenic temperatures is not well documented in the literature, but it can be compared to a similar fluoropolymer commonly used in cryogenic cable insulation, polytetrafluoroethylene (PTFE).  In addition to the material properties, the performance of the installed compound cable can also depend on many factors including the effectiveness of thermal anchoring.

%therefore, the thermal conductivity of isolated FEP is measured here and shown in figure \ref{k_FEP}. 

%\begin{figure}[htbp]
%    \centering
%    \includegraphics[
%        trim=0 0 0 0, clip,
%        width=\columnwidth
%    ]{figures/FEP_k_measurement.png}
%    \caption{Thermal conductivity of FEP.}
%    \label{k_FEP}
%\end{figure}
 
Following a measurement technique similar to \cite{woodcraft2010}, a heater resistor on a copper plate is suspended by the wire samples under test, and the temperature at this plate ($T_\text{high}$) is measured. 
The other ends of the wires are at a constant temperature ($T_\text{low}$), anchored to points in good thermal equilibrium with the 4K plate of the cryostat. 
The input power ($\dot{Q}_\text{in}$) and the stabilizing temperature at the suspended copper plate follow the relationship in Equation \ref{Qin}, where for each material component of the wire $A$ is the cross-sectional area, $\ell$ is the length, and $k(T)$ is its temperature dependent thermal conductivity.

\begin{equation}
    \label{Qin}
    \dot{Q}_\text{in}(T_\text{high}) = \frac{A}{\ell} \int _{T_\text{low}} ^{T_\text{high}} k(T) dT
\end{equation}

\begin{figure}[htbp]
    \centering
    \includegraphics[
        trim=0 0 0 0, clip,
        width=\columnwidth
    ]{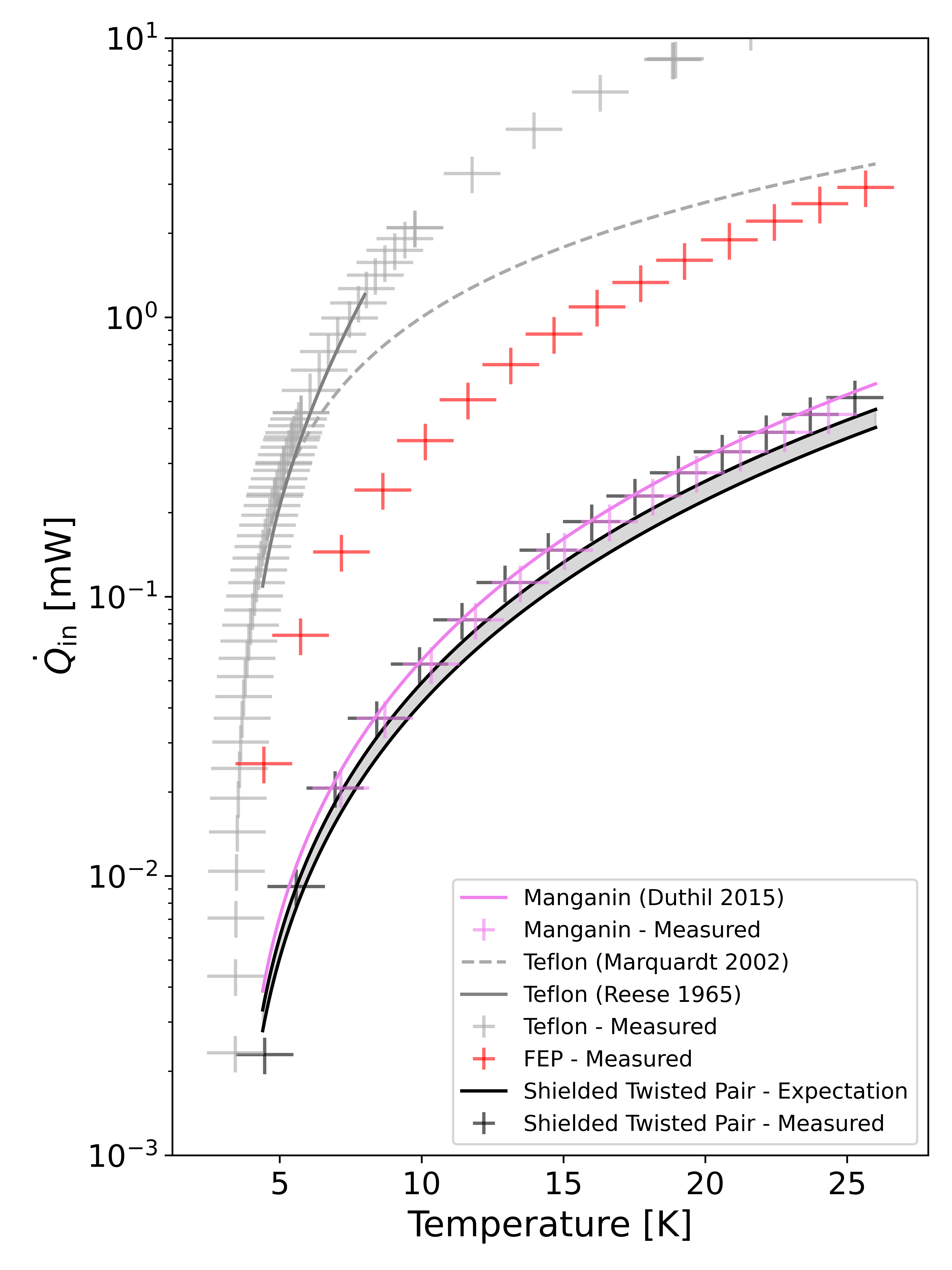}

    \caption{Measurements (crosses) and expectations (lines) based on Equation \ref{Qin} with base temperature at $T_{\text{low}}\approx 3.5$ K and $\ell\approx2.6$ cm. {\bf{Manganin}}: 8 wires of 32 AWG. The expected $\dot{Q}_\text{in}$ is determined by using thermal conductivity of manganin in \cite{duthil2015}. {\bf{Teflon}}: 4 strips with cross-section $\approx 0.01\times0.0024$ m$^2$. The two lines for expected $\dot{Q}_\text{in}$ shown are derived from the compiled thermal conductivity fit in \cite{marquardt2000} (also found in the NIST Cryogenic Material Properties Database) and an extrapolation from thermal conductivity data in \cite{reese1965}, which had measurements closest to our temperature range. {\bf{FEP}}: With 
    %This sounds like the two strips had each twice as much cross section
    no expectation from the literature to compare against, we compared our measurements of one strip of FEP (cross-section $\approx 0.01\times0.0031$ m$^2$) with a measurement of two strips with twice the cross-section, and determined the error based on expecting otherwise identical material properties. The error bars on these measurement data points are derived from comparing these measurements of one and two strips. {\bf{Shielded Twisted Pair}}: Dimensions for the cross-section of these wires are shown in Figure \ref{Thermal}. The expectation is based on the calculated thermal conductivity contribution from FEP and manganin, and it also considers the theoretically expected contribution from SS 304. The shaded region is the range in expected value which includes the lower and upper bound of the conductance contribution from insulation.}
    \label{Thermal2}
\end{figure}

\begin{figure*}[!t]
\centering
\subfloat[]{\includegraphics[width=1.5\columnwidth]{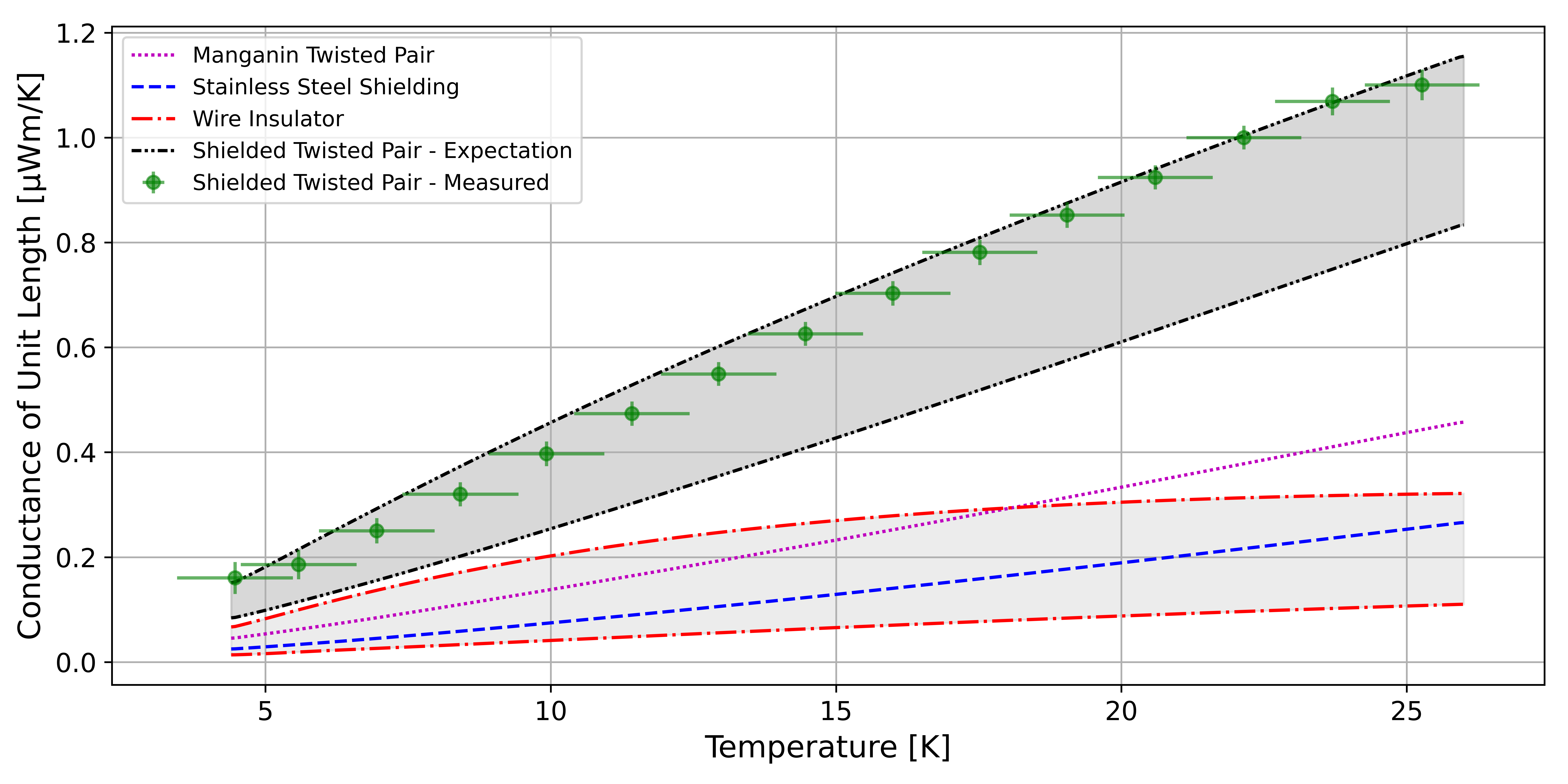}%
\label{figures/Conductance_twp.png}}
\hfil
\subfloat[]{\includegraphics[width=0.5\columnwidth]{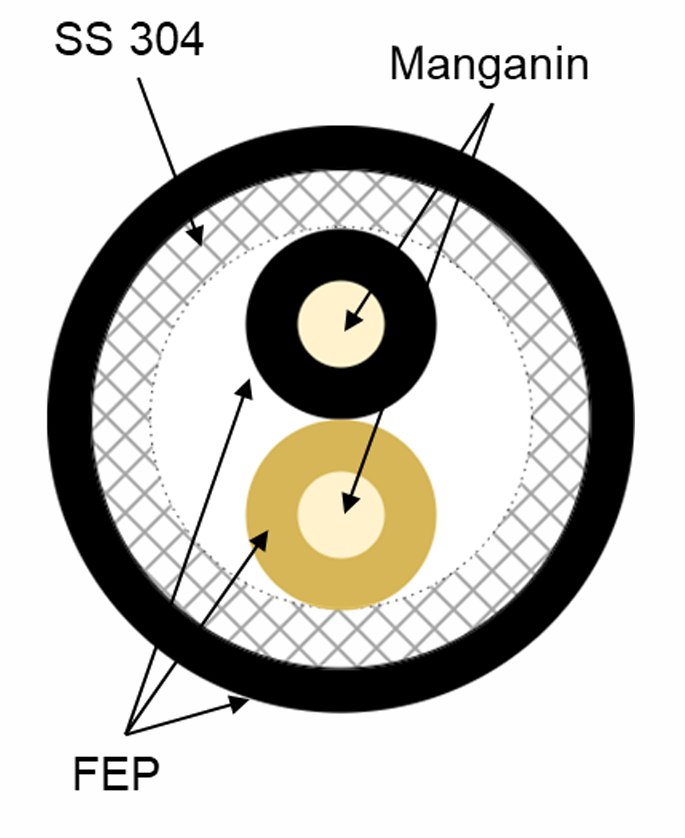}%
\label{fig_second_case}}
\caption{(a) The lines represent the derivative of Equation \ref{Qin}, $\frac{d\dot{Q}_{\text{in}}}{dT_{\text{high}}}$ from data in Figure \ref{Thermal2}, with unit length $\ell = 1$ m and $T_\text{low} = 3.49$ K for the different components of four prototype shielded twisted pair wires. The gray highlights the range between the lower and upper bound estimates of FEP/Teflon and the sum of all the wire components for the four prototype wire samples with cross section shown. The green crosses show measurements and error for conductance and stabilizing temperature ($T_\text{high}$) for the same dimensions of wires. (b) Cross section diagram for shielded twisted pair wires. Manganin wires are 36 AWG with .0045 inch FEP coating wrapped in 44 AWG Soft Bare Type 304 Stainless Steel Braid with 14 wires. The outer FEP coating is 0.004 inch thick. The nominal diameter of the entire wire is 0.044 inches.}
\label{Thermal}
\end{figure*}

For each input power at the heater resistor, $\dot{Q}_\text{in}$, we waited for an estimated dwell time for the system to stabilize at $T_\text{high}$ before stepping to the next value of $\dot{Q}_\text{in}$.
This dwell time was approximately one to ten hours, depending on the sample under test and $\dot{Q}_\text{in}$.
We found the stabilized temperature $T_\text{high}$ and its associated uncertainty for each $\dot{Q}_\text{in}$ step by filtering the data for contiguous segments with both small local derivative and low standard deviation, following the approach in \cite{tsan2021}.
These filtered data were averaged to get the value of $T_\text{high}$.
For a small number of data points with high $\dot{Q}_\text{in}$, the dwell time was slightly underestimated and the system did not reach a stable temperature.
For these points, we applied Newton's law of cooling to fit an exponential model and predict the stabilized temperature $T_\text{high}$.
The difference between this predicted $T_\text{high}$ and the average of the final five
temperature values recorded at that step, along with the uncertainty in the
predicted $T_\text{high}$ from the exponential temperature model, were used to
derive an overall uncertainty for these values.\footnote{
    The specific calculations for this temperature analysis were 
    carried out using an experimental open-source python package developed
    by the authors called cryopy, which may be found at
    \url{https://bitbucket.org/unmcosmology/cryopy}.
}

%We find the stabilizing temperature $T_\text{high}$ and
%its associated uncertainty from averaging over reduced subsets of the data
%for the corresponding value of $\dot{Q}_\text{in}$. We select these subsets based on specific stabilization criteria, %small local derivative and standard deviation, following approaches similar to those used in the literature \cite{tsan2021}. For each range of constant $Q_{in}$, we first filter the data to select contiguous subsets with low local derivatives. From these, we further identify areas of low variability by calculating the local standard deviation at each temperature point, using a neighborhood of five points on either side. We then took the average of these filtered data to get the
%value of $T_\text{high}$. 
%For a small subset of high-temperature steps, the required dwell time was slightly underestimated and the system %did not reach a stabilizing temperature. To address this, we applied Newton’s law of cooling to fit an exponential %model of the copper plate temperature, from which the stabilizing temperature could be predicted. The
%difference between this predicted $T_\text{high}$ and the mean of the final five
%temperature values recorded at that step, along with the uncertainty in the
%predicted $T_\text{high}$ from the exponential temperature model, were used to
%derive an overall uncertainty for these values.

In Figure \ref{Thermal2}, our measurements are shown along with the predictions based on the thermal conductivity from the literature where available \cite{duthil2015, kuster1968, reese1965,marquardt2000}.
Each sample was measured with a similar length and base temperature.
Measurements of manganin wire and the prototype cable showed consistency with expectations, within the measurement uncertainty.
Strips of the insulating materials FEP and Teflon (PTFE) were more challenging to mount with good thermal contact, as these materials were thicker and not very compressible compared to wire samples, and have high thermal contraction at cryogenic temperatures relative to all other materials. Additionally, material thermal properties have shown to vary widely in the literature \cite{marquardt2000}. The measurements of Teflon showed higher than expected thermal conductivity compared to the predicted model for Teflon properties from 4 to 300 Kelvin from the NIST Cryogenic material properties database \cite{marquardt2000}. However, these data are reasonably well matched to an extrapolation from \cite{reese1965}, which characterized Teflon between 1 and 4 Kelvin.

The cryogenic thermal conductivity of FEP is not well documented. 
The thermal conductivity of FEP at room temperature is expected to be lower than Teflon, and our measurements show that it has lower conductivity at cryogenic temperatures as well.
Because we had no literature values to compare to for FEP, we performed two measurements with different amounts of material to help gauge the uncertainty in our technique.
We measured the thermal conductivity of two identical strips of FEP, and then removed one strip and measured again.
The measured points in Figure \ref{Thermal2} have error bars that reflect the difference between the measurement of one strip, multiplied by two to account for the smaller cross-section, and the measurement of two strips. 

Based on these measurements, we show the expected thermal conductivity for the readout wiring in Figure \ref{Thermal}, where the expected and measured thermal conductivity of a single shielded twisted pair is shown along with the contributions from its sub-components.
The contribution from the additional shielding and insulation is more than the thermal conductivity from the manganin conductors alone, more than doubling the total thermal conductivity of the shielded twisted pair cable compared to a bare twisted pair. 

%The lower bound is estimated from dedicated tests of the thermal conductivity of samples of FEP.  The upper bound is a conservative estimate assuming that FEP’s known thermal conductivity at room temperature is constant at cryogenic temperatures.

Additional tests characterizing the performance of the entire integrated cable were also performed, with thermal power from the cable, between 300K and 4K plates, determined by comparing increases in temperature at the cold plates to power vs. temperature load curves taken in an identical cryostat configuration immediately before these cable tests. This included tests checking if the amount of thermal power deposited by the cable onto the cold plate changed whether it was wrapped in low-emissivity materials, which showed no difference. Tests of heat-sinking techniques and their impact on power deposited by the cable were also carried out, which as expected had significant variations depending on the quality of the thermal interface. 
The additional components of the cable including the insulation and shielding do not just contribute additional thermal conductance to the colder cryogenic stages, but also add additional thermal insulating layers that make it more difficult to effectively heat sink the cable.
Initial tests of installation of the cable with non-invasive heat-sinking techniques, clamping sections of the wire with copper and copper foil, resulted in total thermal power at the 4K stage that was over an order of magnitude greater than expected from measurements of its components.

%left bottom right top

%\begin{figure}[htbp]
%    \centering
%    \includegraphics[
%        trim=5 0 45 20, clip,
%        width=\columnwidth
%    ]{figures/Cable_TC_estimate_withFEPMeasuremnt_072725.png}

%    \caption{(Left) The solid lines follow equation \ref{Qin} with $T_{low} = 3.49$ K for the different components of 4 prototype cable wires and the gray highlights the range between the lower and upper bound estimates of FEP and the sum of all the wire components for the wire samples with cross section shown (right) and length ($\ell = 0.02526$ meters). The green crosses show measurements and error for input power ($Q_{in}$) and stabilizing temperature ($T_{high}$) for the same dimensions of cables. Manganin wires are 36 AWG with .0045 inch FEP coating wrapped in 44 AWG Soft Bare Type 304 Stainless Steel Braid with 14 wires. The outer FEP coating is 0.004 inch thick. The nominal diameter of the entire wire is 0.044 inches.}
%    \label{Thermal2}
%\end{figure}

%left bottom right top

%left bottom right top
\subsection{Electrical Testing of Prototype Cold Readout Configurations}

\begin{figure*}[!t]
\centering
\subfloat[]{\includegraphics[trim=40 0 0 0, clip,width=1.25\columnwidth]{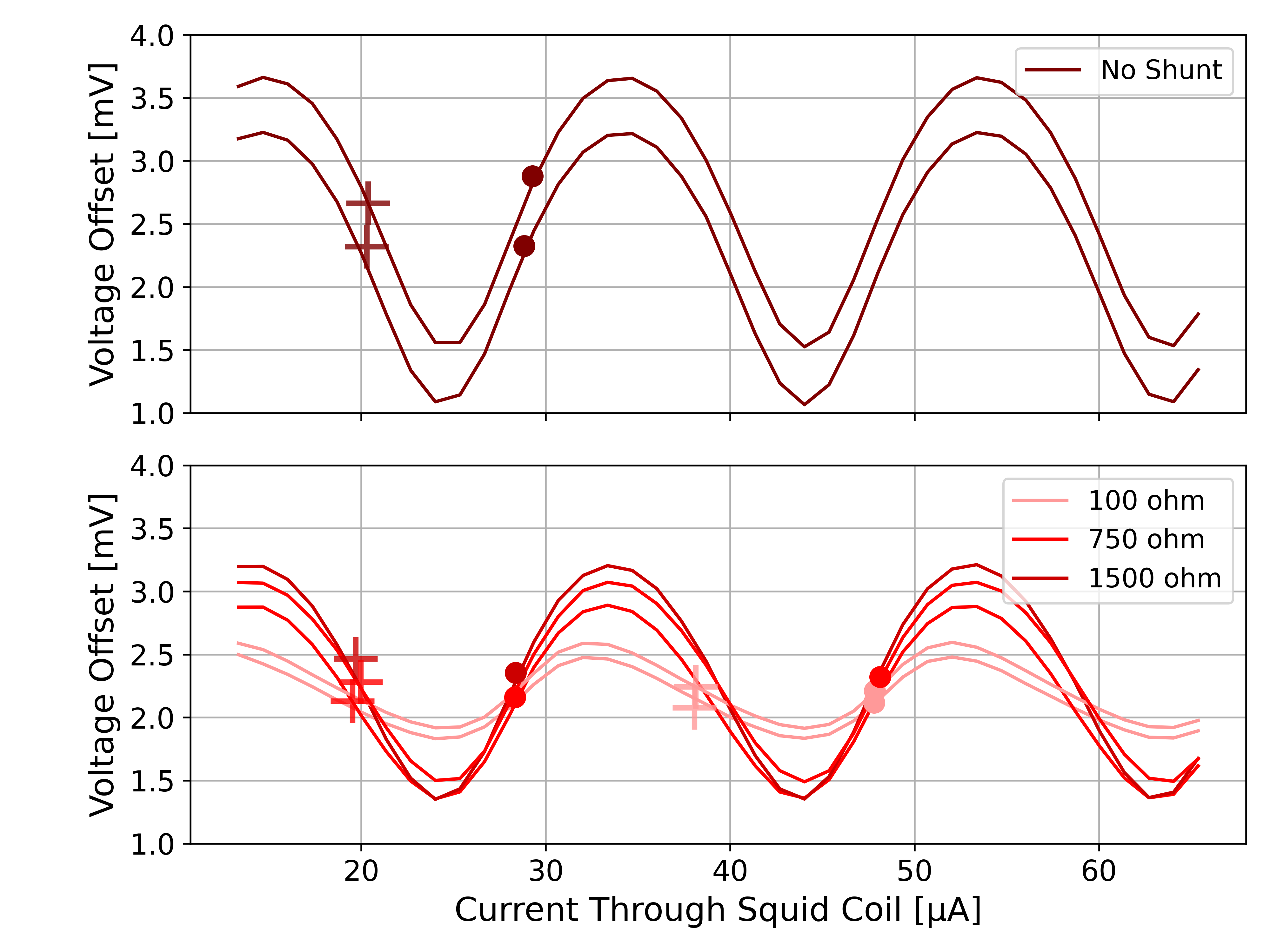}%
\label{fig1}}
\hfil
\subfloat[]{\includegraphics[width=0.75\columnwidth]{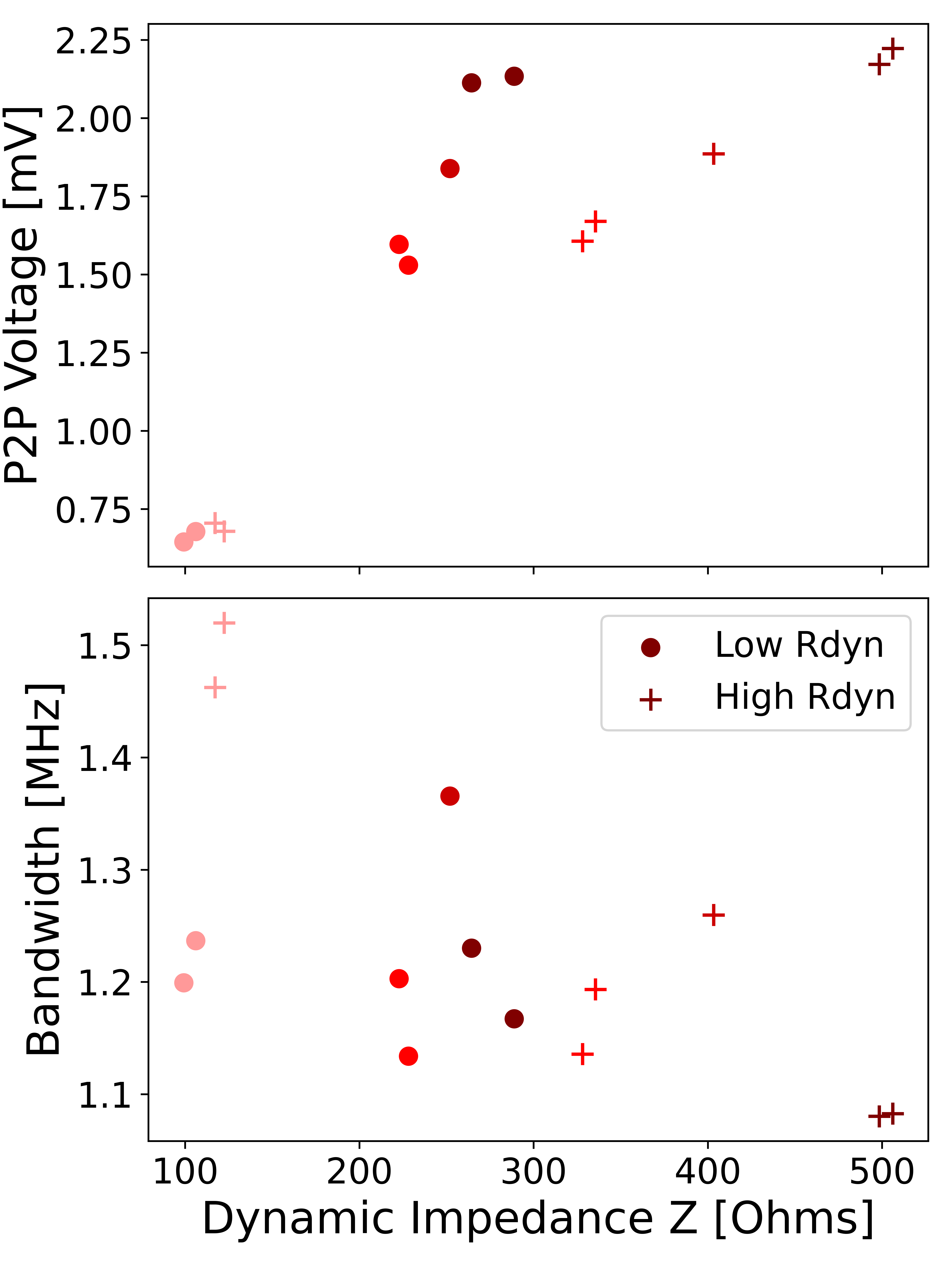}%
\label{fig2}}
%\hfil
%\subfloat[]{\includegraphics[width=2\columnwidth]{figures/Summer24_P2P_Bandwidth_080725.png}%
%\label{fig3}}
\caption{(a) $V$-$\Phi$ curves for SA13s with varying shunt resistances. The cross is the selected tuning point on the downward slope (high $R_{\text{dyn}}$), and the dot is the selected tuning points on the upwards slope (low $R_{\text{dyn}}$). (b) The peak-to-peak voltage (top) and bandwidth (bottom) against dynamic impedance for SA13s with varying shunt resistance for both tuning points.}
\label{Summer24}
\end{figure*}

Dedicated measurements were made to study the effects of different configurations of the prototype cold readout including the SSA and cold wiring. 
This includes tests performed with the dfMux ICE readout system \cite{bandura2016}, a system with a very short cryogenic wiring run to the SSAs ($\approx 6$ in), a wiring configuration complementary to the standard TDM system with MCE electronics and much longer cryogenic wiring. 

The SSAs historically used in TDM readout of TES detectors are a design by NIST that consists of six banks of 64 SQUIDs that can be connected in series or in parallel. %\cite{NIST_SSA_paper}. 
These are connected in two different configurations: three banks in series and two in parallel (3x2), or two in series and three in parallel (2x3). Connecting SQUIDs in series increases gain as well as noise, and connecting SQUIDs in parallel leaves the gain unchanged while noise and dynamic impedance is decreased \cite{max2018,Welty1993}. 
The SSA design used in TDM readout for fielded CMB instruments is the SA13ax in the 3x2 configuration \cite{henderson_advact_2016}.
In the 2x3 cconfiguration, the SA13ax design has demonstrated a bandwidth of $10$ MHz \cite{Doriese2016}.

Another property of the SSA configuration that affects the bandwidth of the cold readout is its output impedance, often referred to as the SSA's dynamic impedance, which is defined as $R_{\text{dyn}} = \frac{dV}{dI_b}$, where $V$ is the SSA voltage offset and $I_b$ is the bias current. Higher SSA dynamic impedance can limit overall system bandwidth. This effect is apparent in Equation \ref{Vin} from \cite{groh2021}, which consolidates the electrical properties of various components into a few terms. The output amplifier voltage is attenuated as dynamic impedance ($Z_\text{out}$) increases.  

\begin{equation}
    \label{Vin}
    \frac{V_\text{amplifier}}{V_\text{SQUID}} = \frac{1}{1+i\omega C_\text{wiring}(Z_\text{out} + R_\text{wiring})}
\end{equation}

A shunt resistor to divert the current across an SSA can alter its properties, such as its dynamic impedance, without the need for a new SSA design \cite{SQUIDshuntpaper,tesche1977}. 
To better isolate and model the effects of shunting SSAs on the bandwidth of this portion of the cold electronics, we characterized the bandwidth with the dfMux ICE readout system, in turn limiting the capacitance effects of the cold wiring on bandwidth. 

%left bottom right top
%%%%%%%%%%%%%%%%%%%5
\begin{figure*}[!t]
\centering
% Row 1: first two subfigures
\subfloat[]{\includegraphics[width=1.29\columnwidth]{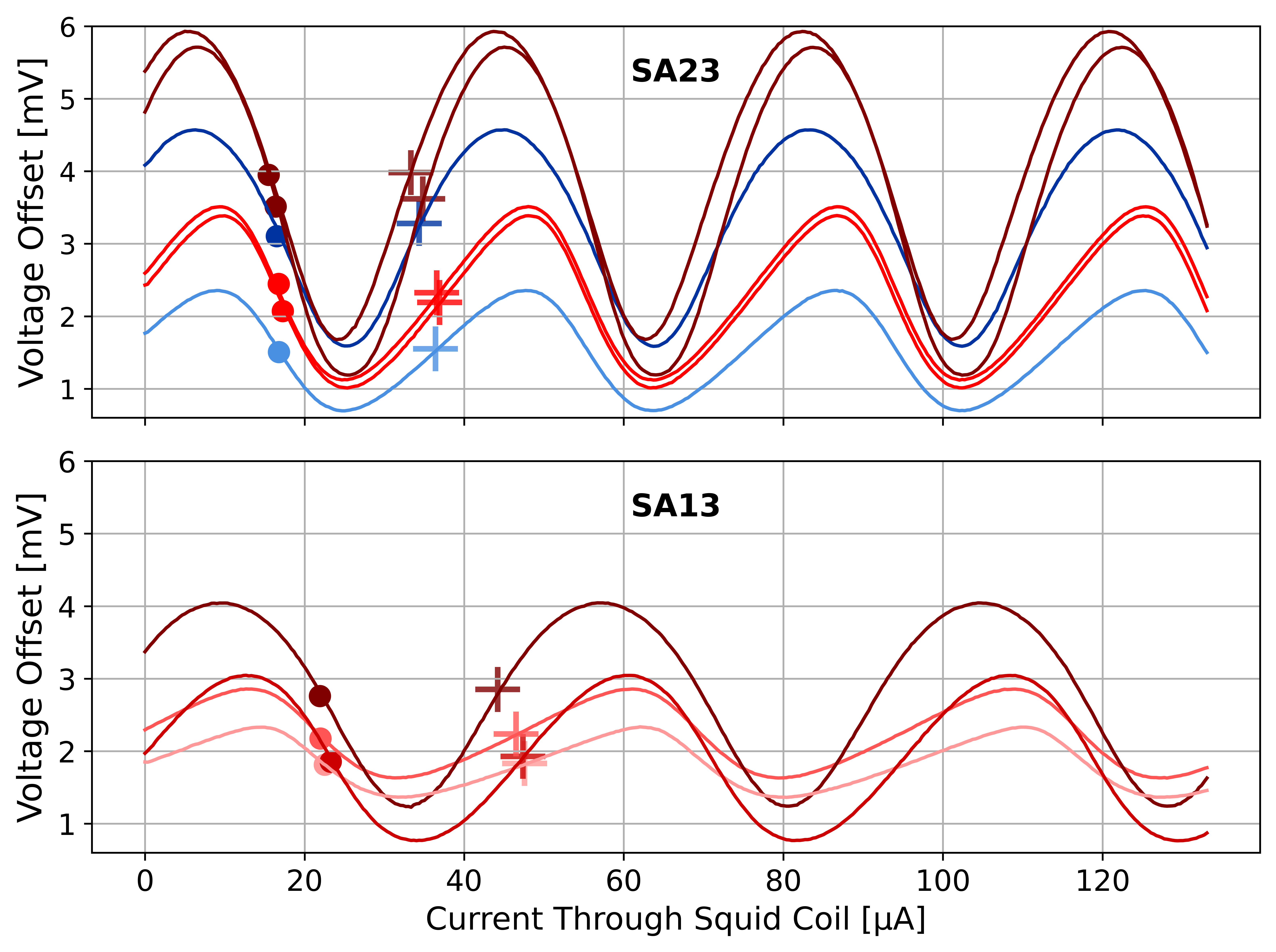}%
\label{fig1}}
\hfil
\subfloat[]{\includegraphics[width=0.71\columnwidth]{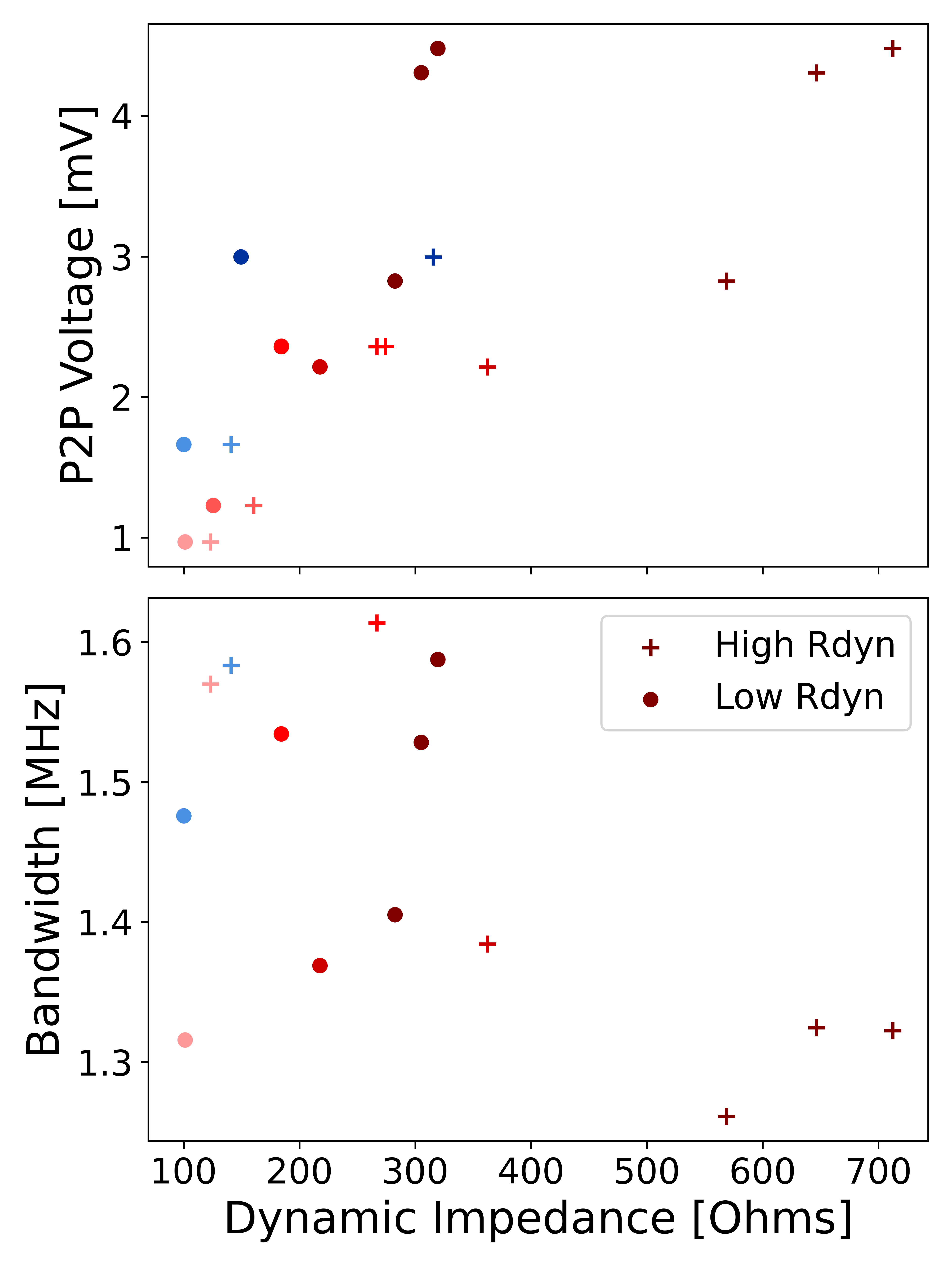}%
\label{fig3}}

\vspace{0.5em} % Space between rows

% Row 2: legend and caption side by side
\begin{minipage}{0.6\columnwidth}
\centering
    \includegraphics[trim=10 30 10 30, clip, 
width =\linewidth]{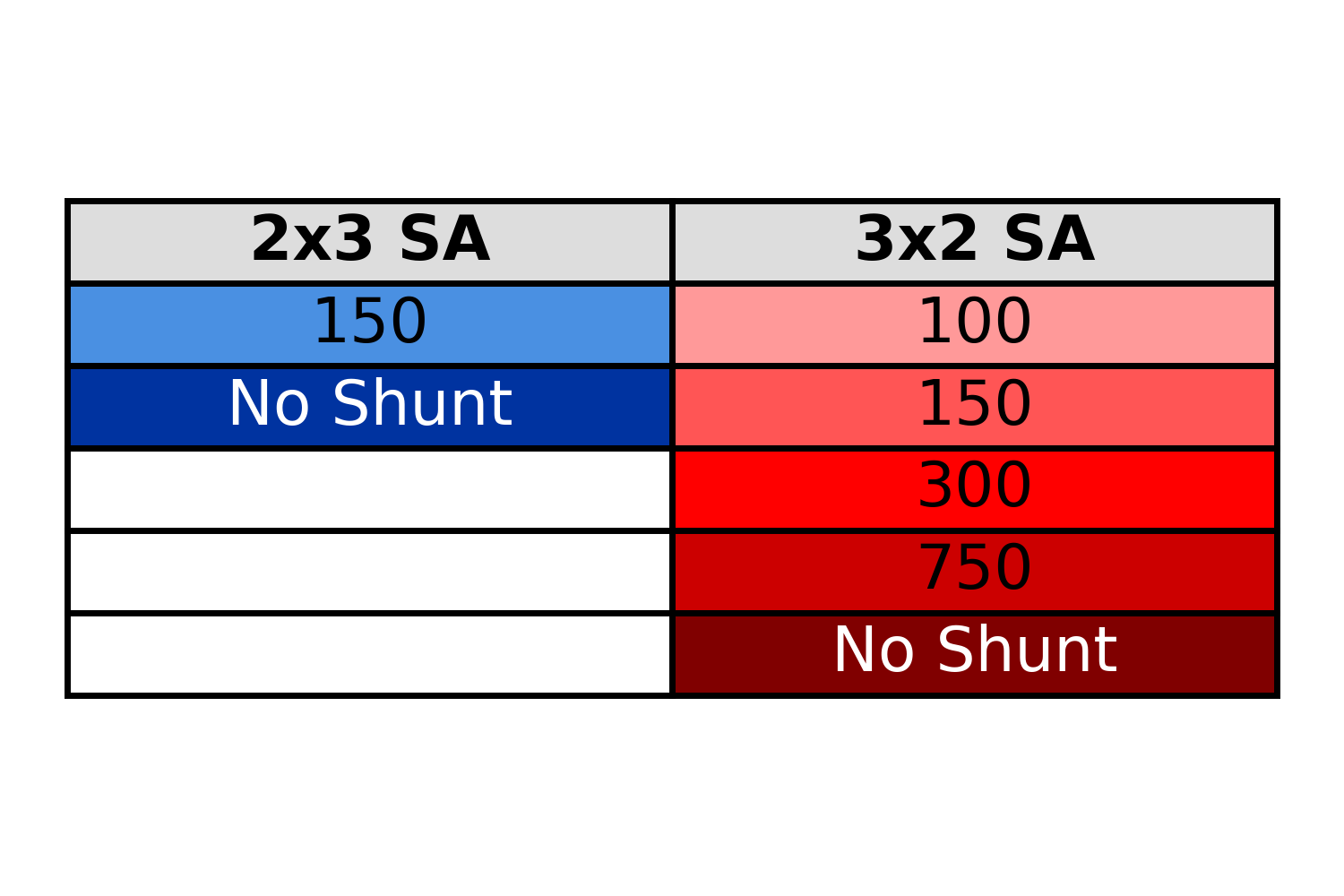}%
    \label{fig2}
\end{minipage}%
\hfill
\begin{minipage}{1.4\columnwidth}
    \caption{(a) $V$-$\Phi$ curves for SA23s and SA13s. The dot is the selected tuning point on the downward slope (low $R_{\text{dyn}}$), and the cross is the selected tuning points on the upwards slope (high $R_{\text{dyn}}$). The legend (left) shows the shunt resistor values for each SSA in their corresponding configuration. (b) The peak-to-peak voltage (top) and bandwidth (bottom) against dynamic impedance for SA13s with varying shunt resistance for both tuning points.}
    \label{Feb24}
\end{minipage}
\end{figure*}

%For each SSA configuration, we measure a detailed set of $V$-$\Phi$ curves, where voltage is recorded for each flux value for a range of bias currents.  
For each SSA configuration, we record the voltage response as a function of flux over a range of bias currents, yielding a detailed set of voltage–flux ($V$–$\Phi$) curves.
The optimal bias current is chosen where the peak-to-peak voltage is maximized, and then operating points are chosen where the slope of the $V$-$\Phi$ curve is greatest.  Once the SSA is tuned, with current bias and flux bias set, we perform a network analysis by sending tones at varying frequencies into the SQUID's input and feedback coils, and measure the voltage response, dividing out the known transfer function of the room-temperature electronics. 
Our estimate of bandwidth of the cold electronics is the frequency value where the signal drops -3dB from the DC value.  

%Preliminary measurements are shown in Figure \ref{Summer24}, showing the modifications to dynamic impedance and peak-to-peak voltage amplitude as the shunt resistance is varied. 
 %Figure \ref{Summer24} shows that preliminary tests demonstrated the effect that we expect:
 
 Figure \ref{Summer24} shows preliminary tests that demonstrate the expected effect for SSAs tuned to the high $R_{dyn}$ working point: lower gain but higher bandwidth from lowering the dynamic impedance of SA13s through shunting.  However, these effects were modest, and the low $R_{dyn}$ working point does not demonstrate a trend in bandwidth at all.
 The competing effects of decreasing gain but higher bandwidth of the SSA must be better understood along with any external limiting factors in bandwidth and performance to fully understand the design space for the CMB-S4 readout design. 
%Measuring the extent of these effects of SSA configuration on overall system bandwidth, gain, and noise will be applicable making design choices for the SSAs that will be used in the finalized CMB-S4 readout system.
Further tests were performed with more variations on the SSA design and configuration, with tests including two versions of the NIST designs - SA13s from 2013 and SA23s from 2023. The SA13 SSAs were only tested in the standard 3x2 configuration, and the newer SA23s were tested in both the 3x2 and 2x3 configurations. However, only two prototype SSAs in the 2x3 configuration were tested, which is a limitation of this work. The SSAs were shunted with different shunt resistance values, or not shunted at all. Figure \ref{Feb24} shows these results, with $V$-$\Phi$ curves of the SSAs marked with the high $R_\text{dyn}$ and low $R_\text{dyn}$ tuning points.
%Figure \ref{rdyn} shows the $R_\text{dyn}$ response overlayed with the  $V$-$\Phi$ curve of an SA23, depicting how tuning the SSA on the negative slope will result in low $R_\text{dyn}$ and tuning on the positive slope results in high $R_\text{dyn}$. 
Higher shunt resistance of the SSA results in higher dynamic impedance and peak-to-peak voltage offset ($V_\text{pp}$), with unshunted SSAs having the highest $R_\text{dyn}$ and $V_\text{pp}$. In Figure \ref{Feb24} this effect is observed for the SSAs in the 3x2 configuration. The SA23s in the 2x3 configuration, due to more SQUIDS being connected in parallel, are expected to have lower dynamic impedance, which can be seen in Figure \ref{Feb24}. 

These results did not show clear trends of the SSA configuration driving the dynamic impedance, peak-to-peak voltage, or bandwidth.  Limitations include the small number of devices tested, but another factor that was not well controlled was the dynamic impedance of the SSA under operation.
The dynamic impedance is not just a property of the SSA design and shunt resistor, but is also a function of flux through the SSA. 
For the NIST SSAs, the design includes two possible working points at the middle of the rising or falling slope of the SSA’s $V$-$\Phi$ response that have significantly different dynamic impedance ($R_\text{dyn}$), as shown in Figures \ref{Summer24} and \ref{Feb24}. It should be noted that in these two tests the wiring to the SSAs had reversed polarity from each other, mirroring the resulting $V_\text{pp}$ curve. In Figure \ref{Summer24} we see that tuning on the negative slope results in high $R_{\text{dyn}}$ and vice-versa, but in Figure \ref{Feb24} we see the opposite relation. 

Another limitation of these measurements with the dfMux ICE readout system was that the SSAs connected to the short cryogenic wiring were at 7K. 
However, this allowed us to compare SSA properties at different temperatures when compared to tests in other systems with similar SSAs at 1 and 4K.
Other tests had shown that the SA23s had a much higher peak-to-peak voltage response than the SA13s (approximately two times higher at 4 Kelvin).
While operating in these higher temperature conditions, the SA23s maintained a high peak-to-peak voltage response, similar to what had been achieved with SA13s at a nominal operating temperature around 4K.%, generally had a higher peak-to-peak voltage response. 

\section{Conclusion}

We characterized the thermal and electrical properties of prototype readout components for CMB-S4 with a goal of understanding the limiting factors for increasing the bandwidth and detector multiplexing rate. These results can inform further integrated tests and system design studies to determine the optimal readout system parameters for CMB-S4, which depend on other system-level factors and constraints.

An accurate model of the thermal conductivity of the cold readout wiring for CMB-S4 is necessary for precisely estimating thermal budgets on the cryogenic stages. 
The prototype readout cables for CMB-S4 included materials without well-documented thermal conductivity at cryogenic temperatures, so dedicated measurements were made of wire samples and isolated materials. 
Thermal conductivity measurements of manganin, FEP, PTFE, and a shielded twisted pair wire are presented here, with measurement uncertainties and comparisons to literature expectations where available. 
Additional tests with the full prototype cable with 50 shielded twisted pairs demonstrated the importance of careful design and testing of heat sinking techniques to fully thermalize such a complex cable, and keep its thermal conductance near expectations.
The expected thermal power dissipated by these readout cables depends on their composition and length, and is part of an overall thermal budget for the cryostat systems. These measurements will help inform options that meet the overall thermal budget.
%Further analysis of cable capacitance and effects on bandwidth will be necessary to determine the viability of this shielded cable for CMB-S4 readout.

%While Manganin and SS 304 have well documented thermal conductivity, FEP does not, making the comparison of experimental data to theoretical expectation difficult. 

%In addition, we present measurements of various SSA designs, determining the parameter space for dynamic impedance, peak-to-peak voltage amplitude, and system bandwidth for different shunt resistances. SA13s in the 3x2 configuration behaved as expected in most respects, even while operating under warm conditions (7K). As the dynamic impedance increased, the peak-to-peak voltage increased and the bandwidth decreased. However, when the tuning point was selected at the low $R_{dyn}$ (positive slope), the effects on the bandwidth were not as clear. The combined measurements with SA13s and SA23s in the 3x2 and 2x3 configurations showed similar results, with the low $R_{dyn}$ (negative slope) bandwidth effects also not very clear. The new SA23 design is shown to have a greater peak-to-peak voltage amplitude than the SA13 operating under similar conditions (7K). 

In addition, we present measurements of various cold readout configurations, characterizing the dynamic impedance, peak-to-peak voltage amplitude, and system bandwidth for different SSA configurations including shunt resistances. SA13s in the 3×2 configuration behaved as expected, even under warm conditions (7K). Decreasing dynamic impedance $R_{dyn}$ reduced the peak-to-peak voltage and appeared to moderately increase bandwidth, but effects at the lowest $R_{dyn}$ tuning points were not conclusive. Combined SA13 and SA23 measurements in 3×2 and 2×3 configurations showed similar trends, with bandwidth effects at low $R_{dyn}$ (negative slope) also unclear. The new SA23 design exhibited a greater peak-to-peak voltage amplitude than the SA13 under comparable 7K conditions.
Further work is needed to better characterize the properties of SA23 devices at typical operating temperatures, with more detailed measurements and monitoring of the dynamic impedance in operation. Further study is also needed for other limiting factors for bandwidth and performance in the system, including cryogenic cables.

\section*{Acknowledgment}

CMB-S4 is supported by the U.S. Department of Energy (DOE), Office of High Energy Physics (HEP) under Contract No. DE–AC02–05CH11231; by the National Energy Research Scientific Computing Center, a DOE Office of Science User Facility under the same contract; and by the Divisions of Physics and Astronomical Sciences and the Office of Polar Programs of the U.S. National Science Foundation under Mid-Scale Research Infrastructure award OPP-1935892. 
Work at SLAC National Accelerator Laboratory was supported by DOE HEP under contract DE-AC02-76SF00515. 
Argonne National Laboratory’s work was supported by the U.S. Department of Energy, Office of High Energy Physics, under contract DE-AC02-06CH11357. 
D.R.B., W.D., and K.H. were supported by DOE HEP under award number DE-SC0021435, and the work at Argonne National Lab was supported by NSF’s Office of Integrative Activities under award OIA-2033199 (an EPSCOR Track-4 Fellowship).
J.P.F. is supported by DOE HEP under award number DE-SC0015655. 
Considerable additional support is provided by the many CMB-S4 collaboration members and their institutions.
The authors thank Matt Dobbs and Joshua Montgomery for developing and continuing to support the dfMux ICE readout system used in these tests.
The authors thank Hiro Jau for his work investigating the sprawling literature on cryogenic material properties.

%The perceived effects of shunting on bandwidth were not as clear when compared to SA13s. 
%The initial test with shunted SA13s tuned on the negative slope behaved as expected. 

%A clear model of the thermal conductivity of the numerous cold wiring is also required. While Manganin and SS 304 have well documented thermal conductivity, FEP does not, making the comparison of experimental data to theoretical expectation difficult. 

%We will redo the bandwidth and spectral analysis testing with SSA shunting at the intended temperature of 4K. Additionally, we will conduct these tests with the intended length of cold wiring and measure its effects of system bandwidth.  


\bibliographystyle{IEEEtran}
\bibliography{IEEEabrv, references}

\end{document}